%% file: oopsla19_main.tex
\documentclass[acmsmall,nonacm]{acmart}

\acmJournal{PACMPL}
\acmVolume{1}
\acmNumber{OOPSLA} 
\acmArticle{1}
\acmYear{2018}
\acmMonth{1}
\acmDOI{} 
\startPage{1}

\setcopyright{none}

\usepackage{booktabs}   
\usepackage{subcaption} 
\usepackage{xcolor}
\usepackage{url}
\usepackage{amsmath}
\usepackage{mathtools}
\usepackage{listings}
\newcommand{\name}{Spacetime}

\lstdefinelanguage{CollectionsP}{%
  language     = Python,
  morekeywords = {pclass, Q, I, P, from, join, select, on, where,
    union, intersect, select, or, and, is, with, as, foreach,
    subset, union, intersect, Join, Project, set, subset, projection, parameter, Getter, GetterSetter, Deleter, Setter, Producer},
}

\begin{document}

\title[GoT]{GoT: Git, but for Objects}         


\author{Rohan Achar}
\affiliation{
  \department{Donald Bren School of ICS}              
  \institution{University of California, Irvine}            
  \city{Irvine}
  \state{CA}
  \postcode{92617}
  \country{USA}                    
}
\email{rachar@ics.uci.edu}          

\author{Cristina V. Lopes}
\affiliation{
  \department{Donald Bren School of ICS}              
  \institution{University of California, Irvine}            
  \city{Irvine}
  \state{CA}
  \postcode{92617}
  \country{USA}                    
}
\email{lopes@ics.uci.edu}          

\input{0_abstract.tex}


\maketitle

\input{0_abstract.tex}
\input{1_intro.tex}

\input{2_anatomy.tex}

\input{4_formal}
\input{5_implementation.tex}
\input{6_benchmarks.tex}
\input{7_related.tex}

\input{8_conclusions.tex}

\bibliography{oopsla19_main}

\appendix
\input{app_A.tex}
\end{document}

%% file: 0_abstract.tex
\begin{abstract}
We look at one important category of distributed applications characterized by the existence of multiple
collaborating, and competing, components sharing mutable, long-lived, replicated objects. The problem
addressed by our work is that of object state synchronization among the components. As an organizing principle for replicated objects, we formally specify the Global Object Tracker (GoT) model, an object-oriented programming model based on causal consistency with application-level conflict resolution strategies, whose elements and interfaces mirror those found in decentralized version control systems: a version graph, working data, diffs, commit, checkout, fetch, push, and merge. We have implemented GoT in a framework called \name
, written in Python. 

In its purest form, GoT is impractical for real systems, because of the unbounded growth of the version graph and because passing diff'ed histories over the network makes remote communication too slow. We present our solution to these problems that adds some constraints to GoT applications, but that makes the model feasible in practice. We present a performance analysis of \name\ for representative workloads, which shows that the additional constraints added to GoT make it not just feasible, but viable for real applications. 

\end{abstract}

%% file: 1_intro.tex
\section{Introduction}
\label{sec:intro}

Distributed computing is the backbone of many large-scale applications seen today, from online
gaming to machine learning. All distributed computing scenarios revolve around a shared computation
state that is worked upon by distributed components. Over the years, several architectural
styles have emerged for distributed applications: client-server, peer-to-peer, map-reduce,
etc. Different architectural styles are suited for different categories of distributed applications. State
synchronization, however, is a common problem that all distributed systems need to address to
some extent; different architectural styles impose different constraints to how to solve it.

\begin{figure}
\centering
\includegraphics[width=0.4\textwidth]{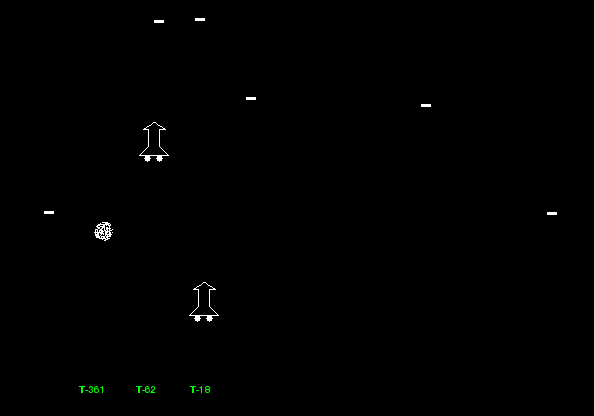}
\hspace{1cm}
\includegraphics[width=0.4\textwidth]{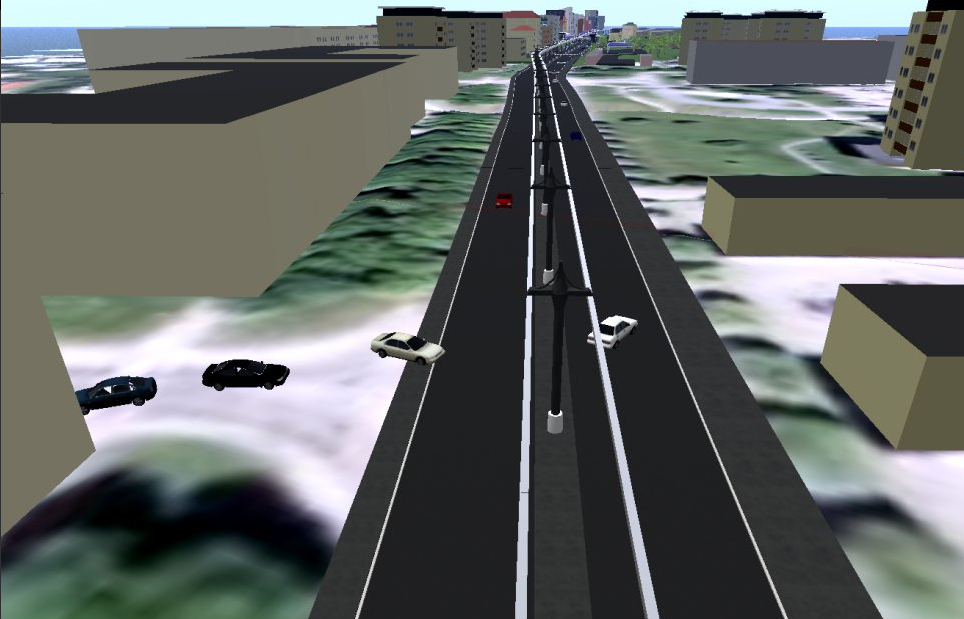}
\caption{Examples of virtual environments for AI-bot competitions. Left: game inspired by the classic Attari's Space Race game. Right: traffic simulation with different fleet operators.}
\label{fig:apps}
\end{figure}

In this work, we are particularly interested in a category of distributed applications characterized by the existence of many components collaborating, and competing, over shared, long-lived, highly mutable state. Examples include multiplayer gaming and distributed multi-agent simulations. In these applications, the components executing or representing agents need to know the state of the world more or less constantly, so the shared state needs to be available locally. When the state of the world changes, those changes may need to propagate to all agents, and inconsistent changes made by different agents need to be resolved. Moreover, the appropriate semantics for resolving inconsistencies is highly dependent on what the logic of the application. Examples of such applications are shown in Figure~\ref{fig:apps}.

This is a well-known category of distributed applications, with many decades of industry attention, especially in the gaming industry and in military R\&D. The main problem that needs to be solved in these applications is data synchronization with as good performance as possible. In practice, this is a hard problem, for many reasons~\cite{CAP}. First, there is the fundamental problem of network latency: a typical 'ping' between the US West Coast and Amsterdam is 130ms, of which 67\% is spent in fiber optic cable at almost the speed of light\footnote{https://wondernetwork.com/pings}. Any attempts to implement strong data consistency will result in severely slowing down the local execution of each component. For that reason, strong consistency is usually not followed; instead, these applications use weaker consistency approaches such as sequential consistency~\cite{lamport:1979}, eventual consistency~\cite{Vogels:2009}, causal consistency~\cite{Ahamad1995}, and others, all of which make dealing with shared state more difficult, but provide higher availability. Second, network delays increase noticeably with the amount of data that is sent across. While compression helps, ultimately the data that needs to be sent depends heavily on the application itself. This knowledge has the perverse effect of increasing the complexity of the applications, as performance-oriented engineers look for application-specific opportunities to reduce the amount of data that is transferred across the network. For example, a server may need to maintain complex information about the state of each client, so to tailor state updates for each of them. Third, there are no general rules for when and how to synchronize the data; the right approach depends on the application. The naive approach of synchronizing every single state change may result in unacceptable application performance. Typically, well-engineered games and simulations buffer changes that are sent only and exactly when needed -- again, typically, at the expense of increased complexity in the code.

Our approach is based on a simple idea: that synchronizing mutable state among distributed components can be modeled as a problem of version control, specifically \emph{decentralized} version control, an idea made popular by tools such as Git. If two components, C$_1$ and C$_2$, have identical copies of the same object O$_1$, and then both change the state of that object locally, both changes are locally valid. Components may decide to clone objects in a specialized heap that we call {\bf dataframes} ("repositories" in Git) and then push/pull the changes, either onto each other or onto a central component, at which point changes will be merged and conflicts resolved -- but even at that point the copies won't necessarily be in the same state. 

We take this simple idea, and formalize a programming model called Global Object Tracker (GoT) that captures a simplified version of Git for in-memory applications objects replicated among multiple  components.\footnote{According to \url{https://github.com/git/git/blob/master/README.md}, one of the several meanings of Git is Global Information Tracker. We re-appropriated that meaning for objects.} In essence, \emph{GoT is the formalization of an object-oriented programming model based on causal consistency with application-level conflict resolution strategies whose elements and interfaces are taken from decentralized version control systems}. GoT is at the core of a framework we have developed called \name
\ that supports distributed components doing sub-second commit, push, and pull operations. \name\ is currently implemented in Python, as that is the main language of the Artificial Intelligence community, but GoT is language independent.

It is not our intention to replicate and formalize the complexity of Git. For example, we do not explore Git branches or capabilities of going "back in time" by checking out older versions while the application is running. The focus of this paper is the precise formalization, implementation, and assessment of the basic operations of decentralized version control on a single, uni-directional branch for purposes of giving a strong organizing principle to replicated distributed objects. To that end, our work makes the following contributions:
\begin{itemize}
    \item We present a new programming model for replicated objects based on a popular decentralized version control system, Git. While Git is widely-known for versioning files, its use in real-time replicated objects manipulated by programs poses a number of challenges, which are discussed and addressed in GoT.
    \item We present a formal specification of this Git-like programming model, GoT. To the best of our knowledge, this is the first time such a model has been formalized.
    \item We identify and solve the challenges of GoT in order to make it feasible in practice: how to deal with potentially very large revision histories.
    \item We demonstrate that GoT can be implemented in practice by presenting one concrete implementation of it, \name; we show that \name\ performs remarkably well under realistic workloads that represent the applications we are currently supporting.
\end{itemize}

Replication package: \name\ is publicly available at \url{https://github.com/Mondego/spacetime}. At the Artifact Evaluation
phase, we will provide a replication package with all the source
code and the setup for replicating the applications and performance experiments reported in this paper.

The rest of the paper is organized as follows. In Section~\ref{sec:anatomy}, we present the programming model informally using a concrete application. Section~\ref{sec:formal} presents the formal specification of GoT without any practical considerations. Section~\ref{sec:implementation} describes the additional constraints we devised that make the model feasible in practice. An experimental assessment of \name\ is given in Section~\ref{sec:expt}, and Section~\ref{sec:conclusions} concludes the paper.

%% file: 2_anatomy.tex
\section{Programming Model: Anatomy of \name\ Applications}
\label{sec:anatomy}

  

\begin{table}
  \begin{tabular}{cl}
    \toprule
    Dataframe Function & Purpose \\
    \midrule
    read & Read one or all objects of atype. \\
    add & Add one or many new objects of a type. \\
    delete & Delete one or all objects of a type. \\
    \midrule
     & Objects are modified directly in code. \\
    \midrule
    commit & Writes changes to object graph. \\
    commit & Get changes from object graph. \\
    \midrule
    push & Writes changes to remote dataframe. \\
    pull & Get changes from remote datafrane. \\
    \bottomrule
  \end{tabular}
  \caption{API table for a dataframe}
  \label{fig:api}
  \end{table}

\begin{figure}
\centering
\includegraphics[width=0.8\textwidth]{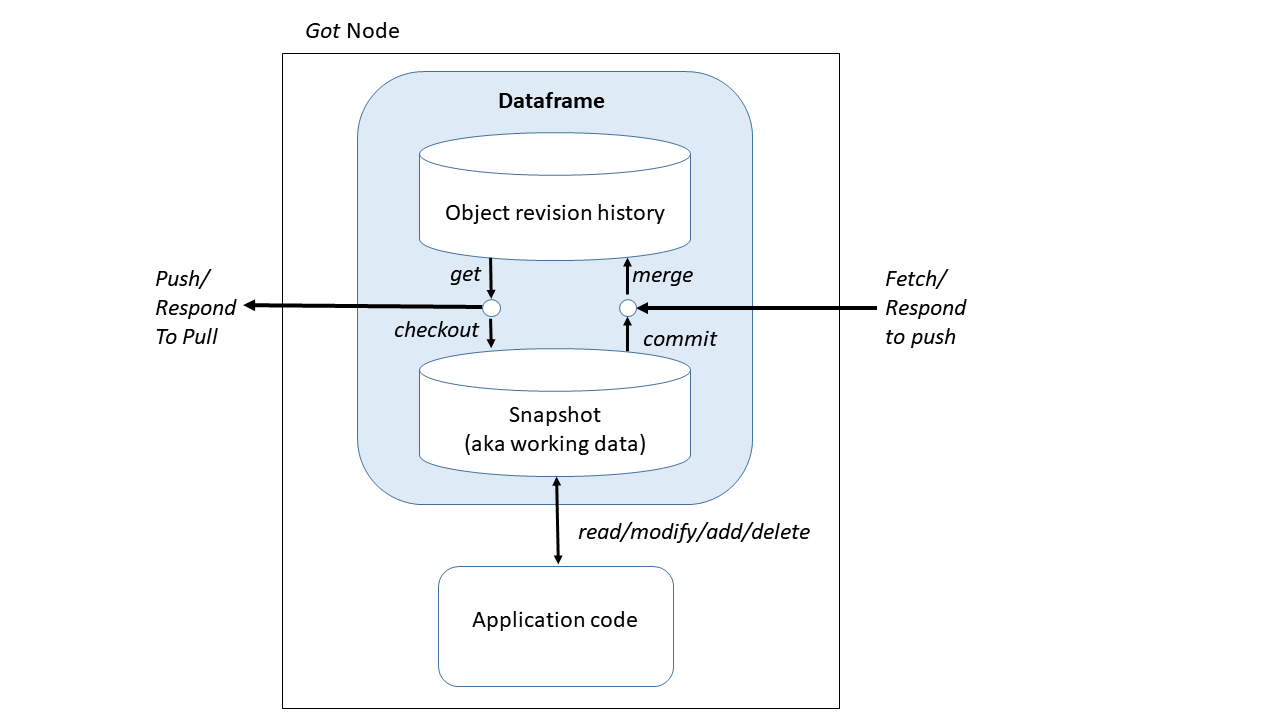}
\caption{Structure of a GoT node. Arrows denote the direction of data flow.}
\label{fig:gotnode}
\end{figure}

Before presenting GoT formally, we begin by explaining what \name\ applications look like using a simple virtual environment example: Attari's-inspired multi-bot Space Race shown on the top-left of in Figure~\ref{fig:apps}. The terminology we use follows that of Git. The only exception is the concept of "repository" which we denote here by {\bf dataframe}. A dataframe is an object heap (separate from that in python) that consists of "checked-out" in-memory objects under version control. The concept is similar to normal, non-bare Git repositories.\footnote{In Git, repositories may be bare or non-bare. Bare repositories contain just the history, but not a copy of the files. We don't have bare dataframes, although those may be introduced in the future if they prove to be useful.} Figure~\ref{fig:gotnode} illustrates the structure and main operations of GoT nodes, which parallels that of Git nodes -- the main difference being that objects, not files, are being tracked. The listed operations -- commit, checkout, fetch, merge, push -- have the same semantics as those operations in Git. Pull, which is used extensively in this section, is the sequence fetch+merge+checkout. An API reference table is provided in Table~\ref{fig:api}

\subsection{Data Model}

The development of \name\ applications starts by identifying the types of objects that will be tracked by the dataframe, and shared under version control. In these applications, "objects" are actual programming language-level objects. As such, they are defined via classes or types. These classes/types need to define which parts of the objects should be tracked, and which parts should not. We have chosen to do this statically and declaratively, as changing the data model at runtime would complicate the data synchronization model unnecessarily. To illustrate this process, Listing~\ref{lst:sr_datamodel} shows the one of the classes of objects that are shared in the Space Race game, along with their most important methods. The rest of the class defintions can be seen in the Appendix~\ref{sec:app_a}.

\begin{lstlisting}[language=Python,basicstyle=\small,
label=lst:sr_datamodel, captionpos=b, caption=The data model for the multiplayer Space Race game.]
@pcc_set
class Player(object):
    oid = primarykey(int); player_id = dimension(str)
    ready = dimension(bool); winner = dimension(bool)
    def __init__(self):
        self.oid = random.randint(0, sys.maxsize)
        ...other initializations, including non-shared fields...
        self.world = World() # Example of non-shared field
    def act(self):
        ... do something smart with the ship...
\end{lstlisting}

The first thing to notice in Listing~\ref{lst:sr_datamodel} are the extra declarations using our language: {\bf @pcc\_set} (decorator of classes), and {\bf primarykey} and {\bf dimension} (special attributes). The declaration {\bf @pcc\_set} indicates that objects of that class are to be tracked. The other declarations define which fields are to be tracked. For example, the Player class defines 5 dimensions that must be tracked, one of which, \texttt{oid}, is the primary key that uniquely identifies the object in the entire universe of objects. All tracked objects can define a primary key field, which must be unique. If a primary key is defined, then objects can be picked up from the dataframe with a reference to the key. If it is not defined, the dataframe assigns a random unique id to the object and the objects can only be retrieved as a part of the collection of objects of that type.

\subsection{GoT Nodes}

\begin{figure}
\centering
\includegraphics[width=0.5\textwidth]{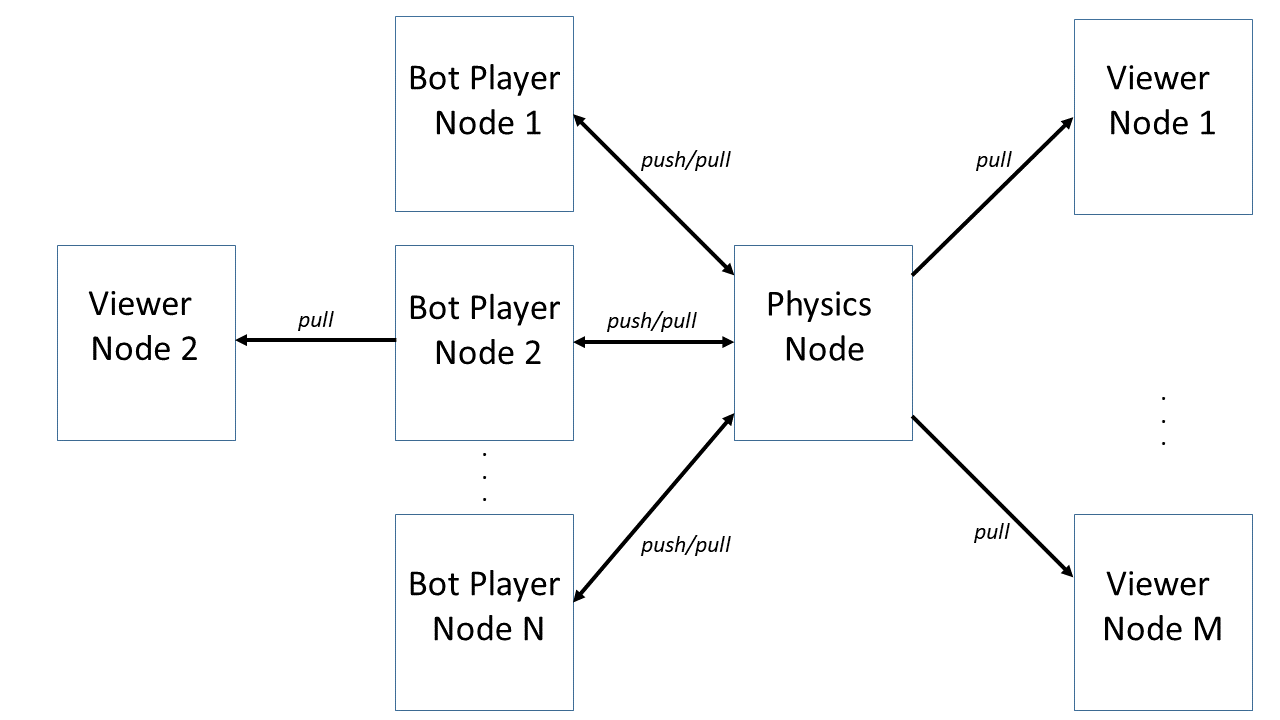}
\caption{Structure of the Space Race distributed application. Each node resides on a separate process/machine/geographic location.}
\label{fig:sr_gotnodes}
\end{figure}

In the case of Space Race, we divided the application into three types of GoT nodes: (1) the world/physics simulator, (2) a simple player bot, and (3) a simple viewer that displays the state of the world. The physics simulator is to be the authoritative component on the majority of the state of the simulated objects, but not all. We explain this in the next subsection. The physics simulator also doubles as the enforcer of the game rules. Figure~\ref{fig:sr_gotnodes} depicts the overall structure of a Space Race deployment. Because there is only one shared world/game, there is only one physics simulator, but there can be any number of player nodes and any number of viewer nodes. There can also be other types of nodes, such as nodes with UI for human players, nodes for compiling statistics, etc., but, for simplicity sake, we do not describe those in this paper.

\begin{lstlisting}[language=Python,basicstyle=\small, numbers=left,
label=lst:sr_physics, captionpos=b, caption=The Physics simulator node.]
class Game(object):
    # Declarations

    def __init__(self, df):
        self.dataframe = df  # store the dataframe
        self.world = self.setup_world() # set up the game.
        self.dataframe.commit() # Push asteroids into the version graph.

    def play(self):
        game_over = False; x = x_gen()
        while not game_over:
            start_t = time.perf_counter()
            self.dataframe.checkout()
            
            for p in self.dataframe.read_all(Player): 
                if p.oid not in self.current_players: # new players
                    self.current_players[p.oid] = p; p.ready = True

            self.move_asteroids(); self.move_ships(); self.detect_collisions();
            self.dataframe.commit()

            elapsed_t = time.perf_counter() - start_t
            time.sleep(Game.DELTA_TIME - elapsed_t)

def sr_physics(dataframe):
    game = Game(dataframe); my_print ("READY FOR NEW GAME")
    while True: 
        game.play(); my_print ("GAME OVER"); time.sleep(WAIT_FOR_START)

def main(port):
    node = GotNode(sr_physics, server_port=port, dataframe="spacerace.got",
                   Types=[Player, Ship, Asteroid])
    node.start()
\end{lstlisting}

Listing~\ref{lst:sr_physics} shows almost the entire code of the Physics node, with just a couple of methods and constants missing. Starting at the bottom, in lines 32--34 we create the GoT node for the physics simulation. \texttt{GotNode}, as the name indicates, is our main class for creating GoT nodes. Its arguments are, from left to right: the entry function to run the node, the port to listen on the network for push and pull requests, the name of the local dataframe, and the types of objects that are to be tracked in the dataframe of this node. These need to be declared as {\bf @pcc\_set} with dimensions to be tracked, as explained in the previous section. The entry function is shown in lines 27--30. These entry functions always receive a dataframe as an argument, so they can perform the data versioning operations on it. In this case, we create a \texttt{Game} class (line 28) that implements the physics simulation as well as other game rules. Several initializations occur in setup\_world (line 6), many of which add objects such as asteroids into the dataframe. After all changes are added to the dataframe, we commit the changes, so that the asteroids will be effectively under version control (line 7). After initialization, the game is played (line 30) via its \texttt{play} method (lines 9--25). The play method consists of a loop that mainly simulates physics, but also has some game functions. At every iteration, it performs a checkout of the objects (line 13), implements the logic (lines 15--21), and commits the changes at the end (line 22). In lines 15--17, it checks if new players have joined the game by checking whether there are new Player objects in the dataframe. If so, it adds them to the game, and lets them play. Lines 19--21 are the main physics functions: asteroids and ships are moved by delta time, and collisions are dealt with. These operations change the shared state of these objects: they end up calling the \texttt{move} methods of Ship and Asteroid that can be seen in Appendix Listing~\ref{lst:sr_datamodel_extended}, which change fields that are under version control. The physics loop runs in 20 fps, so 50ms frames\footnote{"Frame" and "fps" are part of the standard terminology of physics and game engines, as they all operate with these precisely-timed loops explained here. Each precisely-timed iteration is a "frame," not to be confused with our dataframes (although the two concepts are related).} that, in our case, also determine the rate of checkout/commit operations in the Physics node.

Next, we explain the Player nodes. We show only the simplest type of player we have implemented, whose logic is shown in Listing~\ref{lst:sr_datamodel}. When students develop their own AI, they simply write a similar type of node as the one shown here, but with different logic, and, possibly, with different data synchronization strategies. Listing~\ref{lst:sr_bot} shows the most important snippets; the entirety of the code of our player nodes can be seen in Appendix~\ref{sec:app_a}.

\begin{lstlisting}[language=Python,basicstyle=\small, numbers=left,
label=lst:sr_bot, captionpos=b, caption=The Player nodes.]
SYNC_TIME = 0.3 # secs
def bot_driver(dataframe, player_class):
    dataframe.pull()
    my_player = player_class(dataframe)
    dataframe.add_one(Player, my_player)
    dataframe.commit(); dataframe.push()

    my_player.init_world()
    while True:
        start_t = time.perf_counter()

        dataframe.pull()
        survived = my_player.act()
        dataframe.commit(); dataframe.push()
        ... player logic ...
        elapsed_t = time.perf_counter() - start_t
        sleep_t = SYNC_TIME - elapsed_t
        if sleep_t > 0:
            time.sleep(sleep_t)

def main():
    args = ... # parse command line args
    player.start(get_class(args.player))
\end{lstlisting}

One important difference between the Player node and the Physics node presented before is embodied in line 24 of Listing~\ref{lst:sr_bot}. Here, we are binding the player node to a remote dataframe given by the command line argument \texttt{got} (line 24). This argument is expected to be a \texttt{got} URL, of the form \url{got://somehost.edu[:port]/spacerace.got}. This is equivalent to defining the \texttt{remote} origin in Git. In this case, this is supposed to be the URL of the Physics simulator dataframe (see Figure~\ref{fig:sr_gotnodes}). Pull and push operations will be directed to this remote dataframe. Our \texttt{bot\_driver} function (starting in line 2) first pulls all the objects from the remote dataframe (line 3), then creates a Player object and adds it to the dataframe (lines 4--5), and then commits the changes locally and pushes them to the remote dataframe (line 6). This is so that the physics simulator/game receives the new player object when it checks outs its local dataframe (lines 15--17 in Listing~\ref{lst:sr_physics}). From then on, out bot is on a loop of pulling the objects (line 13), doing some local actions on them (line 14), commiting the changes locally and pushing them to the remote dataframe (line 15).  

Our player node is on a timed loop of 300ms (lines 1 and 20), so it only checks the shared state every 300ms. Other player nodes may want to do something different. In any case, this is where network latency plays an important role that cannot be ignored. We are doing push/pull operations to a remote dataframe; depending on the relative locations of the physics node and the player node, each of these operations can take anywhere from 20ms to 200ms or more, depending also on how much data is to be synchronized. As such, the player nodes will always be "behind" the physics node in absolute time, no matter how fast they pull. This is an important aspect of these kinds of applications that has a tremendous influence in their design; the effect will be even more clear in the description of viewer nodes.

Viewer nodes are observers of the simulated world, and their local changes are not supposed to be propagated to other nodes. In our implementation, viewers are graphical components (we use \texttt{pygame}) that display the state of the simulation in a nice, smooth manner. In order to animate the objects, our viewer simulates them at 60fps. This has two benefits: the animation is smooth, and state synchronization becomes simply a matter of compensating for drifting clocks and for acquiring the new objects that are added to the world by other nodes, rather than being the primary means of knowing the state of the world. Rather than trying to be constantly in lock step with the physics simulator, something that would be unreliable, the viewer assumes its role as an autonomous component that operates on its copy of the shared state.\footnote{What we did with our viewer follows the standard practice in the gaming and simulation industries. The difference between our model and the many approaches in use today is our use of version control as the fundamental organizing principle of distributed shared objects.}

\begin{lstlisting}[language=Python,basicstyle=\small, numbers=left,
label=lst:sr_vis, captionpos=b, caption=The Viewer nodes.]
SYNC_TIME = 0.5 # secs
def sync(dataframe, world):
    while True:
        start_t = time.perf_counter()

        dataframe.pull()
        ships = dataframe.read_all(Ship) # Do we have new ships?
        for s in ships:
            if s.oid not in world.ships:
                world.ships[s.oid] = s

        time.sleep(SYNC_TIME - time.perf_counter() + start_t)

def visualize(dataframe):
    dataframe.pull();    world = World()
    for a in dataframe.read_all(Asteroid):
        world.asteroids[a.oid] = a
    
    vis = Visualizer(world)
    threading.Thread(target=sync, args=[dataframe,world]).start()
    vis.run() # Run pygame on the main thread 
    
def main():
    args = ...parse arguments...
    vis_node = GotNode(visualize, dataframe=args.got, Types=[Asteroid,Ship])
    vis_node.start()
\end{lstlisting}

The viewer is the most complex node in the Space Race example, because it executes two threads: the data synchronization thread and the pygame thread. Listing~\ref{lst:sr_vis} shows one of the two parts of the viewer nodes, namely the entry point and the data synchronization thread. Similarly to the player node, the viewer node also binds to the remote dataframe given by the \texttt{got} URL in the command line (line 25). One slight difference is that the viewer is only interested in Ship and Asteroid objects, not in Player objects (line 25). This shows how nodes can tailor these specifications to fit their own needs, with the consequence of lowering the amount of data transferred on pull/push operations. The entry function (lines 14--21) starts by pulling the objects, and storing the pulled asteroid objects in a local dictionary. More importantly, it spawns the data synchronization thread (line 20), and then proceeds to run the \texttt{pygame} visualization object (line 21). At that point, there are two threads: one that is updating the screen at 60fps, changing the state of asteroids and ships along the way (not shown in this listing), and one that pulls data from the remote dataframe (lines 2--12). Note that the \texttt{sync} function (lines 2--17) never commits or pushes; it just pulls.

The \texttt{Visualizer} code is too large to fit in this explanation (see Visualizer class in Appendix~\ref{sec:app_a}); "Sprites" are  basic visual elements in Pygame, and they are associated with one or more images that are then drawn on the screen. Our sprite objects are associated with the shared game objects, namely with instances of \texttt{Asteroid}. Moving these sprites on the screen is done by changing the shared floating point x, y coordinates of the associated game objects, and then setting the integer-based bounding box of the image at those coordinates.

The viewer changes the state of the shared objects only for purposes of animation. However, those changes are never propagated back to the physics node, because the data synchronization thread in Listing~\ref{lst:sr_vis} never commits and never pushes. As mentioned before, this is by design: the viewer is just an observer of the shared state, so when it receives the updated state from the physics server, it simply discards its own estimated values.

\subsection{Conflicts detection and Conflict Resolution}

One of the strong advantages of using a version graph to track object changes is that, detecting conflicts become trivial. A conflict is detected between nodes when revision trees are merged, and a divergence of state versions is detected. This means that different nodes have performed different computations after having read the same state. In the case of the Space Race application, what we have shown so far is free from conflicts. But conflicts would arise, for example, if the AI player would set the position of the ship directly (the Player shown in Listing~\ref{lst:sr_datamodel} sets only the ship's velocity, and that is how all players are supposed to control the ship). This might happen if the player was trying to cheat, but it also might happen accidentally. In that case, there would be conflicts between the player node and the physics node on the \texttt{y} (and even \texttt{x}) fields of the player's ship.

In Git, conflicts are exposed to the users: users have to edit the files and decide whether to keep their own version, or the remote version of the modifications. Clearly, for in-memory real-time application objects, the human-in-the-loop approach would be unfeasible. Instead, we need a mechanism for automatically resolving conflicts. \name\ supports automatic merge conflict resolution by allowing programmers to declare functions specifically for that purpose. For example, additionally to the code seen in Listing~\ref{lst:sr_physics}, the physics node has the following function declared:

\begin{lstlisting}[language=Python,basicstyle=\small, 
label=lst:sr_conflict, captionpos=b, caption=Programmatic conflict resolution for the Physics node.]
def conflict_resolution(conflict_iter, original_snap, my_snap, their_snap):
    for original, yours, theirs in conflict_iter:
        if isinstance(yours, Ship):
            if abs(theirs.velocity) <= World.MAX_SPEED:
                mine.velocity = theirs.velocity
            my_snap.resolve_with(mine)
        else:  # if it is an asteroid
            my_snap.resolve_with(mine)
    return my_snap
\end{lstlisting}

What this means is that upon a merge conflict on any object of class Ship, the physics node uses the velocity set by the conflicting node, if it is under the maximum allowed speed, but keeps its own version of everything else. In this case, this policy makes sense, because the physics node is authoritative for purposes of deciding positions of the objects, but not their velocity -- as long as the velocity honors the rules set by the physics node. In the case of asteroids, no node other than physics is supposed to change any of their parameters; as such, the merge policy simply keeps the physics' node version. For other situations, the merge policies will be different.

An interesting observation is that we can deploy powerful merge strategies. When an unresolvable conflict is detected (and present in the conflict\_iter parameter), a resolution can be made by comparing three different versions of the object. The conflicting versions (mine and theirs), and the version that was the last version that both conflicting versions claim as their predecessor (the original). Mechanisms like version vectors~\cite{versionvector1} employed in industry standard databases, like Riak~\cite{riak} and Apache Cassandra~\cite{cassandra2014apache}, are capable of detecting conflicts, but are not capable of determining the original version from which they deviated. Additionally by versioning the whole state together, and not on a per object basis, semantic conflicts on cross-object relations can also be detected, and resolved. A three way merge with three versions of the whole state (as shown) is possible.  By comparing the conflicting versions and the original version of the state, we provide the application developer a way to maintain implicit relations between objects.

\subsection{Responsibilities of Merging}

In Git, conflicts are not resolved on a push operation. Instead, when there are conflicts, the receiving repository rejects the push request. The user pushing the changes has two options: push the changes in as a new branch, which defers the resolution to a later time, or pull the new changes in the remote repository, resolve the merge locally, and then push the merged state. In the case of \name, neither options are appealing. If we create new branches on conflicts, the nodes keep drifting apart until the deferred merge is invoked. If the state has branched multiple times, it becomes difficult to merge states. If we choose the second option (i.e. reject the push), the sender may never be able to push changes, as the remote node might progress through the states so quickly that the sender is always rejected. For example, a node trying to set the velocity of its Ship might never be able to do it, because the physics simulator keeps committing new values to the Ship's position every 50ms. 

To counter these effects, we allow conflict resolution also on the receiving end of a push request. That's the case shown in Listing~\ref{lst:sr_conflict}. A good consequence of this is that each node has complete control over its version graph and can merge the conflict using logic that they have. For example, the physics node can always decide to accept its changes over the changes pushed by the players, and the visualizer can always decide to accept the changes pulled from the physics node over the changes that it has simulated.

%% file: 4_formal.tex
\section{GoT: Formal Specification}
\label{sec:formal}

We present a formalization of the version control programming model that underlies \name, called GoT (Global Object Tracker). The formal specification serves to unambiguously describe the concepts and operations, independently of any implementation. We were unable to locate any published work with a formal specification of Git. As such, we believe GoT, which, in its current state, is a simplified version of Git, is a valuable step towards that goal. 

Figure~\ref{fig:got} shows the formal specification of GoT. Table~\ref{tab:notation} summarizes the metavariables used in the specification. The specification is divided into 5 parts: (1) the definition of dataframe; (2) the interface that the snapshot exposes to the application code; (3) functions of the version graph; (4) the interface that the version graph exposes to the snapshot;  and (5) the interface that the dataframe exposes to remote dataframes. Figure~\ref{fig:gotnode} (page~\pageref{fig:gotnode}) is a good illustration for understanding the different parts of the formal specification, which are explained next. (A word on notation: the arrows in Figure~\ref{fig:got} denote a change of state in either the snapshot or the version graph). 

\begin{table}
\caption{Metavariables}
\label{tab:notation}
\centering{
\begin{tabular}{l|l}
Primitive metavariable & Meaning\\ \hline
$v$ & Value \\
$d$ & Attribute of an object.\\
$t$ & A concrete type. $t: \{d_1, d_2,\dots,d_n\}$\\
$o$ & An object. $o: <t, id(o), z>$\\
$id(o)$ & The unique primary key of the object. $id(o): z[d_{p}], d_{p} \in t$\\
$z$ & State of an object. $z: \{\forall d\in t, d\mapsto v\}$\\
$r$ & An unique identifier for a version. \\
\end{tabular}
}
\end{table}

\begin{figure}
\noindent
\textbf{D1. Dataframe:}

\noindent
\begin{equation*}
\begin{aligned}[l|l|l]
\text{(Dataframe)\ } & D: & <T, G, S> \\
\text{(Types)\ } & T: & \{t_1, t_2, \dots, t_n\} \\
\text{(Version Graph)\ } & G: & <r_{h}, V, E, P> \\
\text{(Vertices)\ } & V: & \{r_1, r_2,\dots, r_n\} \\
\text{(Edges)\ } & E: & \{e_1, e_2, \dots, e_n\} \\
\text{(Single edge)\ } & e: & \delta_{1\rightarrow 2}, \delta_{2\rightarrow 3}, \dots, \delta_{m-1\rightarrow m} \\
\text{(Snapshot)\ } & S: & <r_s, s, \delta> \\
\text{(Objects in Snapshot)\ } & s: & \{o_1, o_2, \dots, o_n\} \\
\text{(Changes in Snapshot)\ } & \delta: & \{id(o)\mapsto\{d\mapsto v\}[new|mod|del] \} \\
\text{(Remote Nodes)\ } & P: & \{p\mapsto r_p\} \\
\end{aligned}
\end{equation*}

\noindent
\textbf{D2. Snapshot Interface:} 

\noindent
\begin{equation*}
\begin{aligned}[l|l|l]
\text{(New object)\ } & S(r_s, s, \delta) \xrightarrow{S.create(o)} & S(r_s, s, \delta \cup \{id(o) \mapsto z\}[new]) \\
\text{(Delete object)\ } & S(r_s, s, \delta) \xrightarrow{S.delete(o)} & S(r_s, s, \delta \cup \{id(o)\mapsto \emptyset\}[del]) \\
\text{(Write attribute)\ } & S(r_s, s, \delta) \xrightarrow{o.write(d, v)} & S(r_s, s, \delta \cup \{id(o)\mapsto\{d \mapsto v\}\}[mod]) \\
\text{(Read attribute)\ } & o.read(d): &
    \begin{cases}
        \delta(id(o), d) & (id(o), d) \in \delta \\
        s(id(o), d) & (id(o), d) \in s \wedge (id(o), d) \notin \delta
    \end{cases} \\
\end{aligned}
\end{equation*}

\noindent
\textbf{D3. Version Graph Functions:} 

\noindent
\begin{equation*}
\begin{aligned}
\text{(Retrieve changes since $r$)\ } & & \\
G.get(r):   & \{e_r, e_{r+1}, \dots, e_{h-1}, e_h\} &\\
        & & \\
\text{(Receive changes: $r_a \rightarrow r_b$)\ } & & \\
G(r_h, V, E) & \xrightarrow{G.put(r_a, r_b, \{e_a, e_{a+1}, \dots, e_b\})} &
        \begin{cases}
            \begin{aligned}
            G(& r_b, V \cup \{r_{a+1}, \dots, r_b\}, \\
            & E\ \cup \{e_a, e_{a+1}, \dots, e_b\})\end{aligned} & r_a = r_{h}\\ 
            Resolve(G, r_a, r_b, \{e_a, e_{a+1}, \dots, e_b\}) & r_a \neq r_{h}\\ 
        \end{cases} \\
        & & \\
\text{(Conflict Resolution)\ } & & \\
G(r_h, V, E) & \xrightarrow{Resolve(G, r_a, r_b, \{e_a, \dots, e_b\})} &
        \begin{aligned}
            G( r_m, V \cup & \{r_{a+1}, \dots, r_{b-1}, r_b, r_m\},\\
             E\ \cup & \{e_a, \dots, e_b, (\delta_{a\rightarrow h} - \delta_{a\rightarrow b}) \cup \delta_{res}, \\
             & (\delta_{a\rightarrow b} - \delta_{a\rightarrow h}) \cup \delta_{res}\})\\
        \end{aligned}
\end{aligned}
\end{equation*}

\noindent
\textbf{D4. Interaction between Snapshot and Version Graph:}

\noindent
\begin{equation*}
\begin{aligned}[l|r|l]
\text{(Checkout)\ } & S(r_s, s, \delta), G(r_h, V, E) \xrightarrow{D.checkout(G, S)} & S(r_h, s \cup G.get(r_s), \delta), G(r_h, V, E) \\
\text{(Commit)\ } & S(r_s, s, \delta), G(r_h, V, E) \xrightarrow{D.commit(S, G)} & S(r_n, s \cup \delta, \emptyset), G.put(r_{s}, r_{n}, \delta) \\
\end{aligned}
\end{equation*}

\noindent
\textbf{D5. Interaction with Remote Dataframes:} 

\noindent
\begin{equation*}
\begin{aligned}[l|r|l]
\text{(Push)\ } & \begin{aligned}
        D(T, G_l, S_l, P), \\
        D(T, G_p, S_p, \emptyset) \\
    \end{aligned}  \xrightarrow{D.push_to(p \in P)} & 
    \begin{aligned}
        D(T, G_l, S_l, r_{lh}), \\
        D(T, G_p.put(P[p], r_{lh}, G_l.get(P[p])), S_p, \emptyset) \\
    \end{aligned} \\
& & & \\
\text{(Pull)\ } & \begin{aligned}
        D(T, G_l, S_l, P), \\
        D(T, G_p, S_p, \emptyset) \\
    \end{aligned} \xrightarrow{D.pull_from(p \in P)} & 
    \begin{aligned}
        D(T, G_l.put(P[p], r_{ph}, G_p.get(P[p])), S_l, r_{ph}), \\
        D(T, G_p, S_p, \emptyset) \\
    \end{aligned} \\
\end{aligned}
\end{equation*}

\caption{The Global Object Tracker (GoT) formal specification.}
\label{fig:got}
\end{figure}

\subsection{Dataframe}

The core of our model is centered around the component called the dataframe. The dataframe is a specialized shared object heap that the application code in every node in GoT interacts with. Each node in GoT gets its own dataframe to manage the state of its objects\footnote{For simplicity sake, since there is only one dataframe per node, we use these two words interchangeably throughout the paper.}. As an object heap, the dataframe must satisfy two important constraints: read stability, and deterministic state update. In the heap of a single threaded programming languages, these constraints are easy to achieve, as the only way to change the state is from the sequential execution of the application code. In parallel and distributed computing, this is harder to achieve because changes can be made from outside the application. To better isolate the effects of external changes, the heap is divided into two components: a snapshot that provides read stability, and a version graph that detects and manages concurrent changes. The snapshot is the local copy of the objects that the application code in the node interacts with. The version graph keeps track of the published history of shared objects. It can receive both changes committed locally by the node, and changes made externally by external nodes. Changes that made to the version graph do not affect the snapshot without an explicit request from the application code.

Formally, a dataframe $D$ is defined as a tuple $<T, G, S>$: a list of types $T$ declared in a shared data model, a snapshot $S$, and a version graph $G$ to track versions of the objects (Figure~\ref{fig:got} Part D1). 

\subsection{Data Model and Types}
All nodes share the same data model that declares a set of types (denoted by $T$). All objects shared between the nodes must be instances of one of these types. A type, denoted as $t$ is declared with many dimensions $d$. An object $o$ of type $t$ is represented by the tuple $<t, id(o), z>$ (see Table~\ref{tab:notation}). $z$ represents the state table mapping each dimension $d$ to a value $v$. Each object has an identifier $id(o)$ such that the pair $<t, id(o)>$ is globally unique. Example types, Player, Ship, and Asteroid, used in a Space Race game, are shown and described in Listing~\ref{lst:sr_datamodel}.

\subsection{Snapshot}
The role of the snapshot in the dataframe is to provide the application code with a copy of the shared objects whose updates are under the control of the code. As is typical for an object heap in any language, the application code can read objects, create new objects, delete existing objects, and modify attributes of objects (Figure~\ref{fig:got} Part D2). All these reads and writes are executed upon the snapshot. All writes that are made (updates, additions, deletions) are staged in the snapshot (in $\delta$), ready to be bundled as a diff and applied to the version control graph on a commit. It can also request changes (only when requested by the application code using the checkout primitive) from the version graph and apply the diff received to update the local object state. These two primitives are explained in more details below.

Formally, we define the snapshot to be the tuple $<\delta, s, r_s>$. All writes made to the snapshot are staged in $\delta$. These deltas consist of newly added and modified objects as their state deltas (the whole state is a state delta for new objects), and the identifier for all objects that were deleted. $s$ represents the local state of the objects. $r_s$ indicates that the snapshot was last copied from the version graph at version $r_s$.

\subsection{Object Version Graph}
The version graph $G$ is a Directed Acyclic Graph (DAG) denoted as the tuple $<r_h, V, E, P>$. Each vertex in ($V$) represents a version of the state of all the objects. It is never instantiated in the version graph and is only labelled by a version identifier $r$. Each edge in $E$ is the part of the graph that is instantiated and is represented as $e: \delta_{a\rightarrow b}, e\in E$. It represents the diff of changes ($\delta$) required to move the state of the all the objects from the previous version ($r_a$) to the next ($r_b)$. The graph terminates at the head version denoted by $r_h$, which can be considered the latest state version of all the objects. The version graph can receive object changes through diffs either from the local snapshot or from remote dataframes. The version graph can also receive requests for updates from the local snapshot or the remote dataframe. The version graph responds to these requests with diffs representing these updates. Finally, the version graph also maintains a map of the last know states ($r_p$) of each external dataframe ($p$) that it pushes changes to ($P: p \mapsto r_p$). This allows the version graph to generate the correct diff during a push. 

\begin{figure}
\begin{subfigure}[b]{0.3\linewidth}
\centering
\includegraphics[width=0.8\textwidth]{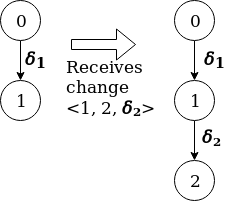}
\caption{Direct increment of version graph.}
\label{fig:normal_continue}
\end{subfigure}
\begin{subfigure}[b]{0.4\linewidth}
\centering
\includegraphics[width=0.8\textwidth]{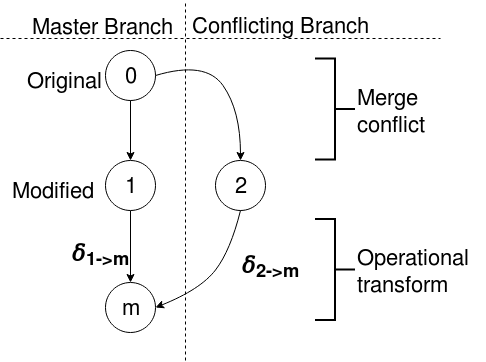}
\caption{Conflicting increment of version graph.}
\label{fig:conf_resolve}
\end{subfigure}
\caption{Merging changes.}
\label{fig:recv_changes}
\end{figure}

The version graph exposes two functions: retrieve changes, and receive changes (Figure~\ref{fig:got} Part D3), which are explained next.

\subsubsection{Retrieving changes}
Changes can be retrieved by specifying a version identifier $r_n \in V$. This version identifier denotes the last version of the state known by the requester. All the edges (diffs) starting from $r_n$ up until the head version $r_h$ are retrieved and returned in order. Applying these diffs, in order, on the state at version $r_n$ would bring the state to version $r_h$.

\subsubsection{Receiving changes}

When receiving changes, the version graph receives an ordered set of edges, $\{e_a,e_{a+1}\dots,e_b\}$, and version identifiers, $r_a$, $r_b$. The edges represent all delta changes, that, when applied to a state at version $r_a$, transform it to the state at version $r_b$.

When applying these changes, the version graph grows. If it receives change from the head version ($e: \delta_{a\rightarrow b}, r_a = r_h$), the changes are not in conflict, and the edges can simply be appended to the existing list of edges in the graph (Figure~\ref{fig:normal_continue}). What this means is that all the changes received were created having first read the latest change in the graph. $V$ is also updated with all the version identifiers that the graph can reconstruct from the delta changes it has. The new head version changes to $r_b$.

However, if the version graph receives a change $e: \delta_{a\rightarrow b}$ with $r_a \neq r_h$ and $r_a \in V$, i.e. a delta that moves the state from version ($r_a$) that is not at the head of the graph to a divergent state ($r_b$), then the changes received are in conflict with all the changes between $r_a$ and $r_h$ already present in the version graph, and must be resolved (Figure~\ref{fig:conf_resolve}). The graph receiving the changes is always responsible for resolving any conflicts.

\subsection{Semantics of conflict resolution}
In GoT, the developer can write a custom merge function to resolve cases of conflict. Conflicts are resolved asynchronously and not under the control of the application code except to supply the merge function. The merge function takes in four parameters: an iterator over objects that are in conflict, and three snapshot states representing the node state at the point of fork, and at the end of the diverging paths. This provides rich semantics for merge write-write conflicts and maintaining cross-object semantics after resolution. Changes are all recorded in the corresponded dataframes. The delta changes in the dataframe returned by the function is picked up as the resolved delta $\delta_{res}$. Listing~\ref{lst:sr_conflict} gives an example of such a merge function. In addition, any number of default merge strategies can be employed by the programmer to automatically merge all conflicts without having to define their own custom function.

With a final resolved state ($r_m$) known as the result of the three way merge, the version graph can be updated. GoT employs a variation of Operational transformation~\cite{Randolph12} (OT) to bring the diverging states to the resolved state. In traditional OT, two functions must be created that when applied on the divergent states, transforms them into the common resolved state. Different data types would need to be handled differently making the creation of these functions extremely difficult. In GoT, we do not create different transformation functions. We create different delta changes that are applied to the divergent states using the exact same function and bring them to an identical state. While transformation functions can be hard to generate, generating the delta required to bring one state to another is simpler. We create a delta containing all object dimensions that are in the later state but not present, or different in the former. In the formal model, we represent the delta from the head version ($r_h$) to the merged version ($r_m$) as  $\delta_{h \rightarrow m} = \delta_{a \rightarrow b} - \delta_{a \rightarrow h} \cup \delta_{res}$. This includes all changes present in the conflicting delta ($\delta_{a \rightarrow b}$) that are not present in (and thus in conflict with) the master branch ($\delta_{a \rightarrow h}$). This difference is represented as $\delta_{a \rightarrow b} - \delta_{a \rightarrow h}$. In addition, we add the delta changes that were obtained from the three-way merge function ($\delta_{res}$). Similarly, the delta from the conflicting version $r_b$ to the merged version $r_m$ is represented as $\delta_{b \rightarrow m} = \delta_{a \rightarrow h} - \delta_{a \rightarrow b} \cup \delta_{res}$. This gives us three new edges that are added to the graph as a result of the operation: the conflicting change ($\delta_{a \rightarrow b}$) that was received, the change from the current head of the graph to the conflict resolved version ($<\delta_{h \rightarrow m}$), and the change from the conflicting version to the conflict resolved version ($\delta_{b \rightarrow m}$).

Any cross-object semantic preserving changes made in the custom merge function would be part of the resolved delta ($\delta_{res}$) and, therefore, would be preserved in the final changes.

\subsection{Interaction between the Snapshot and the Version Graph}

In addition to the standard object heap operations, (defined in Figure~\ref{fig:got} Part D2 and explained above), the snapshot also define two primitives that allow the Nodes to move data between the snapshot and the version graph: Commit, and Checkout (Figure~\ref{fig:got} Part D4). When a commit is invoked, a new globally unique version identifier $r_n$ is created. All the staged changes that are present in $\delta$ represent the change from the last checked out version of the snapshot $r_s$ to the newly created identifier $r_n$. The version graph receives this change resolving conflicts if any. The staging $\delta$ is reset to a null set when the changes are confirmed by the version graph, and the version identifier is updated to $r_n$.

Moving data the other way, the snapshot can be updated with the latest changes in the version graph by executing the checkout primitive. In a checkout, a single delta is created. It is obtained by merging all deltas since the last snapshot version, in order. The delta change is applied to $s$ bring the state of $s$ to the version $r_h$. Checkout and Commit are under the control of the developer and can be called at a rate that meets the requirements of the task performed by the node.

\subsection{Interaction with Remote Dataframes}

With data in the dataframe present as a chain of delta changes; sending, receiving, and requesting changes to other dataframes is simplified. Dataframes provide the application code with two primitives that allow the node to retrieve and receive data from other dataframes explicitly: pull, and push, respectively (Figure~\ref{fig:got} Part D5). A dataframe that wants to initiate requests to another dataframe must first register the remote dataframe. Each remote dataframe is stored as an address ($p$) mapped to the last version known for that node ($r_p$). 

When a node performs a push, it picks up the ordered set of delta changes since $r_p$ from the local version graph and sends it across the network to the remote repository. The remote repository receives the update and grows the version graph, resolving conflicts if it needs to (as defined in (Figure~\ref{fig:got} Part D3). If the remote repository does not have the start version $r_p$ (in case of restarted nodes), the remote repository rejects the push. A full synchronization of the version graph can help recovery in such a situation.

When a node requests a pull from a remote dataframe, it sends the last known version, $r_p$ along with the request so that it need only receive the deltas since that chagne. The dataframe receiving the request, tries to retrieve changes since the requested version (as defined in Figure~\ref{fig:got} Part D5) up to its head version, and returns the ordered set of delta changes. The node making the request then receives the changes and adds it to the version graph, resolving conflicts if any. If the requested version does not exist in the remote node (in case of restarted nodes) the request fails. A full synchronization of the version graph can help recovery in such a situation.

 These two primitives provide the object heap of each Node the ability to share patches of changes between each other. Additional variants of these primitives (pull\_await, push\_and\_wait) can be implemented that offer optimization in the network traffic required for synchronization.
 
\subsection{Consistency Model}
The consistency model supported by GoT is causal consistency. To be causally consistent, four session guarantees must be met~\cite{terry1994session}: read your own writes, monotonic reads, writes follow reads, and monotonic writes.

{\bf Read your own writes.} All writes performed by the node are first written to the snapshot. The snapshot will preserve the write, up until it gets explicitly deleted by either the application itself, or by a conflict resolution at the version graph, both of which are writes that follow the first write. There are no rollbacks allowed in the GoT version graph. {\bf Monotonic reads.} Monotonic reads ensure that once a process observes a state, it cannot observe any state that preceded it. Since the version graph does not allow roll backs, this condition is met. {\bf Write follows reads.} All writes to the version graph must specify the version that was observed on which the changes were executed upon. The state DAG is built using this relation. This ensures that writes always follow reads. {\bf Monotonic writes.} When the same process performs a commit to a version graph, the state of the process is considered to be at the state after the execution of the write. Any successive writes would need to specify either the state at the end of the last write or a state that follows it. This ensures monotonic writes.

%% file: 5_implementation.tex
\section{Implementation: Making GoT Feasible}
\label{sec:implementation}
The formal model, GoT, as presented in the previous section would be impractical when implemented directly for tracking replicated live objects manipulated by programs. In this section, we present many important implementation decisions that make GoT feasible for tracking versions of application objects. Our implementation of GoT, called \name, implements these optimizations.

\subsection{Management of the Version Graph}
Over the course of continuous execution, the version graphs in each node, if implemented true to GoT, grow in size at an unbearable rate. This can very quickly lead to memory exhaustion. For software version control systems like Git, this is relatively insignificant, because the rate at which the graph grows is significantly slower. In \name\ applications, however, the version graphs are updated much more frequently, making the size of the version graph a significant challenge. For example, the physics node shown in Listing~\ref{lst:sr_physics} is doing 20 commits per second! To address the problem, a ledger is maintained per dataframe to record the last update sent to all remote dataframes. Unnecessary deltas are garbage collected when they have been successfully pushed to all other dataframes. 

A similar optimization is employed to compress the version graph in \name. An additional reference counting map is maintained in each node that maps each external node it has interacted with (either by receiving a push/pull request from it, or by sending a push/pull request to it), to the last version of the graph that the external node has acknowledged. On a successful push transaction, the sender updates the reference count, mapping the receiver to the appropriate version. The receiver also updates its reference count, mapping the sender to the version that it just received. With this map, a successful garbage collector can be devised to compress the version graph.

The graph optimizer passes through every version recorded in the graph from the latest version, to the root, for each object. Every version not referenced in the mentioned map, is deleted, and a new edge (with merged diffs) is created to join the parent of the deleted node to the child of the deleted node. If the node belongs to a conflicting branch, the branch is deleted. If the node has multiple parents or children (due to the resolution of a merge conflict), the node is not deleted until all connected branches are deleted.

\begin{figure}
\centering
  \includegraphics[width=0.7\linewidth]{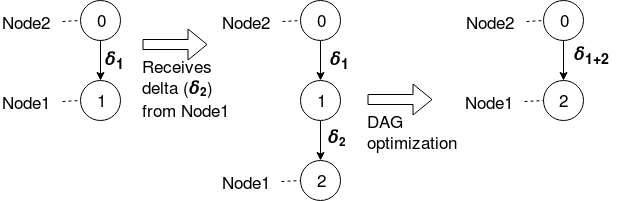}
  \caption{Optimization when receiving simple changes.}
  \label{fig:delta1}
\end{figure}

Figure~\ref{fig:delta1} denotes the map update, and graph optimization, that occurs when a node, say Node0, receives data from another node Node1 (either as a push request from Node1 or as Node1's response to a pull request). In the initial state of the example, the version graph contains a transition 0 $\rightarrow$ 1, through the update $\delta_1$. Node0 knows that after the last interaction with Node1, the state in Node1 was at state version 1. Similarly, the state in Node2 is known to be at state version 0. When Node0 receives a new update from Node1 ($\delta_2$), the version graph is updated, and the state in Node1 is now known to be at state version 2. When node Node2, makes a pull request to Node0, it receives a diff that is the merge of $\delta_1$ and $\delta_2$. Since there are no known nodes at state version 1, this state is deleted. The graph is shortened to two states: 0 and 2. The diff associated with the edge is also updated to include the merged diff $\delta_{1+2}$. When Node2 makes a pull request, it will receive $\delta_{1+2}$, and the last known version of Node2 will be updated to 2. If Node1 makes a pull request, it receives no diff as it is already at the latest. Any new node that makes a pull request receives all the data from the root. No node can make a request from an intermediate version if it has not made at least one prior request from the root version. When the version graph is forked due to a merge conflict resolution, the root of the fork and the join node of the fork are not cleaned up until the fork branch has been deleted.

\begin{figure}
\centering
  \includegraphics[width=0.7\linewidth]{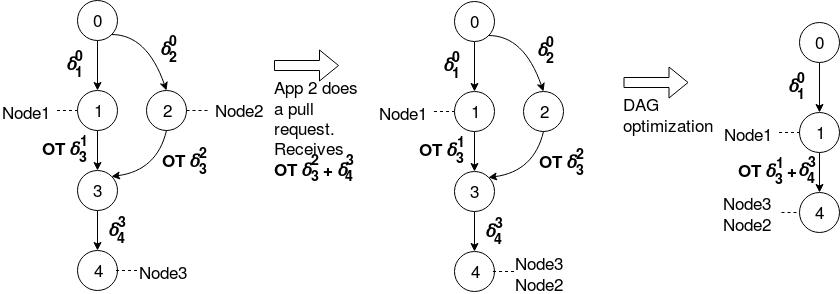}
  \caption{Optimization when sending changes.}
  \label{fig:delta3}
\end{figure}

The version graphs are also optimized in the node that is sending data. To demonstrate, let us look at the optimization that is performed in Figure~\ref{fig:delta3}. The initial state of three external nodes, Node1, Node2, and Node3, are known to the node, say Node0, whose version graph is being optimized. Node2 had a merge conflict resolved in the last push that it performed. After the resolution, Node3 pulled the latest changes, and pushed in a new diff. Node2 is now making a pull request. The node Node0, bundles up $OT \delta_3^2$, that brings the divergent Node2 to the master branch, and $\delta_4^3$ together; and sends it to Node2. The last known version for Node2 is updated in Node0 as state version 4. During the optimization, state version 2 is deleted along with the divergent path $0 \rightarrow 2 \rightarrow 3$. State version 3 is neither marked, nor does it have multiple parents or children. So the state is deleted, and a connection is made from $1 \rightarrow 4$ using the diff $OT \delta_3^1 + \delta_4^3$.

With a simple reference counting garbage collector, the number of diffs that each node needs to maintain in the DAG is in the order of the number of nodes with which it has interacted. This significantly reduces the memory footprint, but it forces the node to hold some state of the other nodes. The tradeoff is well worth it: this is one of the most important decisions that makes GoT feasible in practice.

\subsection{Network Efficiency}

According to GoT (Figure~\ref{fig:got}, Part 5), when nodes push and pull changes from other nodes, they receive changes as a collection of edges starting from the version they were at, up until the latest version in the remote node. The advantage is that any repository can be the remote repository. The entire chain of history is stored at all locations. The same advantage exists when implemented in \name. Nodes can pull from any other node provided the nodes have at least the same version. However, the network cost incurred in sending complete chain of history is significant. If these deltas were merged (squash update in git), it would be significantly smaller. We use this optimization in \name by combining the deltas at the sender's location, and transmitting only one compressed edge that represents the state change. This gives us significant gains in reducing the communication overhead. The tradeoff, however, is that nodes cannot request delta updates from a node they have not visited before. The first pull from any node has to be from the root version history. The trade off is acceptable, as the topology of \name\ applications is usually decided before hand, and nodes do not typically hop between multiple remote nodes.

\subsection{Data Partitioning}
One of the most important requirements to reducing the communication overhead between nodes in is that nodes should only receive data that they need. In distributed computing, very often, different components of an application are doing different tasks, requiring different subsets of the data. Having nodes synchronize over the entire data state implies transferring data to nodes that are never going to use it. For example, in the space race applications, the visualizer never needs objects of the type Player. All it requires to render the game are the objects of Ship and Asteroid. The Git model, and by extension, GoT, does not directly support synchronization of subsets of the data. This is because version tracking becomes harder when the less than the whole state is labelled against a single version. If the repository is broken up, then each part can have a different version progression, and the complexity starts increasing. If every object had its own version graph, then the cost of creating a delta for transmission would be at the order of the number of objects, but every node can choose and synchronize the objects they are interested in. If the the entire state had its own version graph, the deltas are already calculated and ready to be transmitted, but the nodes must receive a replica of the entire state. 

Resolving this tradeoff requires a compromise between partitioning the data into chunks that can be versioned, and not having each node have to subscribe to a large number of such versioned chunks. In \name, we version by type. Objects of the same type are grouped together into collections and synchronized with other nodes that are interested in the same type. The nodes declare the set of types that they are interested in and have one version graph for each type. Updates are composed by iterating over the version graph for each type. Changes that are received are divided by type and distributed to their specific version graph. Since the number of types do not change and are not typically not in the same order of magnitude as the number of objects, the computational efficiency that is traded is not a lot. Another advantage of versioning by types, instead of objects, is that types can never be deleted or added dynamically, whereas objects can be. There are lesser states of change to reason over with named collections. The disadvantage, however, is that dynamic collections are difficult. For example, if there were many ships in the Space Race environment and only a small subset of these objects were actually active, the visualizer might want to subscribe to only those ships that are active. Having to synchronize the state of all ships might be inefficient. Overcoming the difficulty of effective data partitioning while retaining the speed of version graphs is planned for the future.

%% file: 6_benchmarks.tex
\section{Experimental Results}
\label{sec:expt}

\subsection{Experimental setup}

\subsubsection{Workload}

We used the Space Race game that has been extensively described before in the paper. It consists of three types of nodes: Bot nodes, a Physics node, and Viewer nodes. The three types of nodes are arranged in a topology similar to that shown in Figure~\ref{fig:sr_gotnodes}. There are three types of objects that are synchronized: Ship, Player, and Asteroid. The definitions of these types is shown in Listing~\ref{lst:sr_datamodel}. We summarize here the version control operations that each node is programmed to perform, and how often:

\textbf{Space Race bot:} The bot is declared as synchronizing on all of the three types. In the application code, the bot creates an object of the Player class, commits the changes and pushes it to the Physics node. It pulls any changes from the physics node once every 0.3 secs. At receiving a signal from the node signifying the start of the game, the bot enters a loop where it performs an action, commits and pushes the action to the physics node, and receives changes back. The duration of the loop is fixed to take 0.3 secs. During this span, the bot makes one external push, one external pull, one checkout, and one commit.

\textbf{Space Race viewer:} The viewer is declared as synchronizing on only Asteroid and Ship. The viewer also runs in a loop fixed at 0.3 secs, where it just pulls changes from the physics node and renders the view in a parallel thread, using pygame. In each loop, the viewer performs one pull from the physics node, and one checkout to the local snapshot. In our experiments, we used two variations of the viewer, one where the predictions that the viewer makes are checked into the version graph and resolved against the changes pulled from the physics node, and one where the predictions are strictly local and replaced with the physics copy on checkout. The pull does not conflict in the second case.

\textbf{Space Race physics:} The physics node contains the authoritative state of the application. It receives changes from the bots, and responds to requests from both the bots and the viewers. The game runs at a rate of 20 frames per second. The values of the ships and the asteroids are updated in each frame, and committed to the version graph. Changes sent in from the bots are also read in each frame and are used to determine the next state. In each loop (0.05 secs), the physics node is performing one checkout, and one commit. There are no network operations in the game loop. 

\subsubsection{Experiment 1: Latency}

In one set of experiments, we use one physics node, one viewer, and one bot, to measure the performance of version control operations under different conditions.

The independent variables that we varied were the size of the state tracked in the version graphs at each node, and the existence (or not) of functions to resolve conflicts between the changes made by the viewer and the physics node. To vary the size of the state, we varied the number of asteroids (20, 100, and 200 asteroids). Asteroids are synchronized by both the bot and the viewer. Since the viewer predicts the movements of asteroids and pulls changes to asteroids from the server, we set up a merge function in the viewer to merge incoming conflicts. In the cases where we did not want to see a merge, we did not commit the changes made by the viewer into its version graph. 

In each experiment, we measure the median elapsed time to execute six functions: push, fetch, pull (fetch + merge), receive request (both push and pull), commit, and checkout. The former four all include network code in the form of socket reads and writes. The last two methods are local. 

\subsubsection{Experiment 2: Memory}
In a second set of experiments, we demonstrate the effectiveness and the aggressiveness of the garbage collector. The independent variables here are the number of nodes in the system: one bot, one visualizer, and one physics node; and ten bots, one visualizer, and one physics node. We record the number of versions maintained at the authoritative node (the physics node) over time after each operation on the version graph. 

\subsubsection{Hardware and Network Conditions}

We used two machines to conduct the experiments. One machine was an Amazon EC2 instance located in Virginia, US, running Ubuntu 18.04 with kernel 4.15.0-1031 with 1GB of RAM, and one core of an Intel Xeon CPU E5-2676 v3 processor. We used this machine to run the physics node.
The second machine, used for launching the bots and the viewer, was a desktop machine running Ubuntu 16.04.1 with kernel 4.15.0-43-generic with 32GB of RAM, and an Intel Core i7-5820K CPU with 12 cores (with hyperthreading). This second machine was located in the West Coast of the US. 

Ping time between the two locations is typically 62ms, on average. The time reported by the \texttt{hping} tool (\url{http://www.hping.org/}) at the time of the experiments was 72ms. Unlike ping, \texttt{hping} uses TCP, so it may be a more accurate baseline for our experiments than ping.

\subsection{Results}

\subsubsection{Experiment 1}

\begin{figure}
\centering
\includegraphics[width=\textwidth]{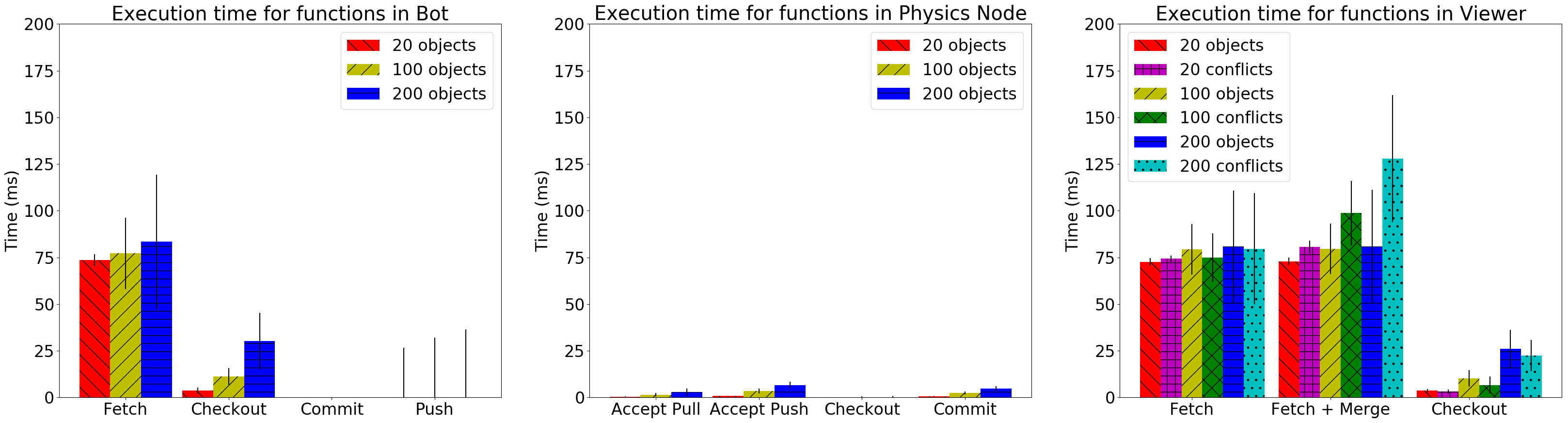}
\caption{Median execution times of multiple functions in each node.}
\label{fig:median_exe}
\end{figure}

Figure~\ref{fig:median_exe} summarizes the results of our latency measurements in experiment 1. From left to right, we show the time it took to execute the several GOT functions at the bot node, the physics node, and the viewer node, respectively. As expected, operations over the network (fetch, push, and pull) reflect the unavoidable network latency of about 72ms.

The first observation to make is that the size of the state does not affect the latency of fetch operations in \name significantly. As we can see from Figure~\ref{fig:median_exe}, the median fetch latency for both the Bot node and the Viewer node is under 75ms for 20 asteroids and 80-83ms for 200 asteroids. 
This is because all communication between the nodes is carried out over deltas only. The small increase in latency is due to seralization/deserialization of more deltas.

A fetch+merge operation, as shown in the viewer bar graph, is composed of both the fetch request and a request to merge the received changes into the version graph. Even though the size of the deltas is larger with more number of objects, we do not iterate over the objects in the delta during the merge. They are simply attached to the version graph. As such, the latency of fetch+merge is practically the same as the latency of fetch -- as long as there are no merge conflicts. The cost of merging deltas into the state is seen only when the bot or the viewer performs a checkout and moves the changes to the snapshot. As we can see, the cost of checkout is significantly higher with increased number of objects. 
However, fetch+merge requests that include conflict resolution are considerably more costly. This can be seen the viewer for those experiments where the objects are conflicted. Moreover, the higher the number of objects that have conflicts, the costlier the process. 

The median elapsed time for push at the bot is close to zero. The bot in our experiments pushes objects in the beginning, and then keeps moving at the same velocity. Since no dimensions are being set during that time, the version graph does not change, and so there are no deltas required to be pushed. Communication does not happen in \name\ if it is not required. On the side of the physics node, the cost of responding to push or pull requests is very low. Most of the time spent in both the push and pull requests is spent in the network.

\subsubsection{Experiment 2}

\begin{figure}
\centering
\includegraphics[width=0.4\textwidth]{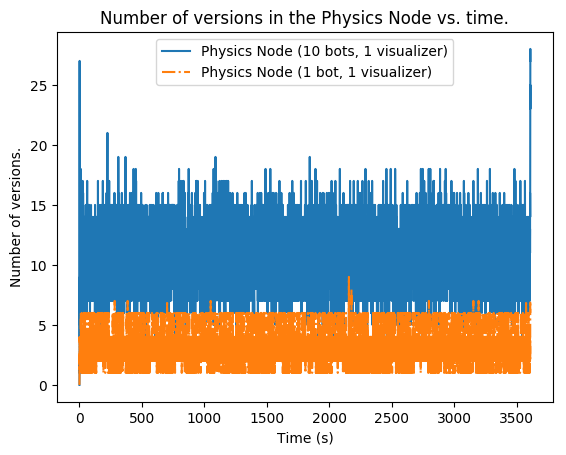}
\caption{The number of revisions that are maintained at each version graph.}
\label{fig:version_graph_node_count}
\end{figure}

Figure~\ref{fig:version_graph_node_count} shows the number of versions maintained in the version graph of the physics node. In particular, we are interested in seeing if the aggressive reference counting garbage collector is effective in curbing the version explosion that can occur in the version graphs. As we can see, the number of versions is not growing over time. This implies that there are no memory leaks in the logic of the garbage collector. When there are multiple nodes communicating with the same node, the number of versions tracked increases as different nodes are likely to be at different states. Write-write conflict resolutions also contribute to an increase in the number of versions. The number of versions in a node is proportional to the number of nodes communicating with it.

%% file: 7_related.tex
\section{Related Work}
\label{sec:relwork}

The problem tackled in this paper is as old as the field of distributed computing itself. We narrow down our focus to the most recent work that relates to ours. But we stand on the shoulders of a large body of work in consistency models~\cite{Lamport1979,Ahamad1995,Tanenbaum2006}, Distributed Version Control Systems~\cite{Rochkind77,Hunt76,Hunt96,Milewski97}, and Operational Transformation~\cite{Ellis:1989,Carlsson:1993,Nichols:1995,Ressel:1996,Randolph12}. \\

\noindent
{\bf Versioning}

\noindent
There have been several distributed systems that use some form of versioning, or history, to detect and resolve conflicts of distributed updates to replicated data. For example, Version Vectors~\cite{versionvector1} are used in database systems, such as Riak~\cite{riak}, to detect conflicts by maintaining a count of updates for each distributed component and passing along this map to each component along with the requested data. These methods do not scale with increasing number of components, as the vectors reach unmanageable sizes. Additionally, any merge resolution strategy only has access to the conflicting versions and not the parent version (because it's a vector, not a graph). Some variations such as Hash Histories~\cite{hashhistories} provide more capabilities to determine the parent version for conflict resolution. 

Most of the work in this area is done in the context of replicated databases, and aims at making the problem of data consistency hidden from application programmers. That is the opposite of our goal. Many optimizations in distributed applications are application-specific, and therefore it is very difficult for these replicated databases to be able to support the variety of needs of these applications. For example, no online multiuser games rely on database replication mechanisms for sharing the real-time game state with the players. Our goal is to provide a simple programming model for data synchronization with which application programmers can have control over what gets synchronized, with whom, and when.

\noindent
{\bf Conflict-Free Replicated Data Types}

\noindent
Conflict-free Replicated Data Types (CRDT)~\cite{shapiro11} are an interesting middle ground between consistency models and programming models. The observation underlying CRDT, which was already visible in~\cite{Oster06}, and became even more evident in Treedoc~\cite{Preguica:2009}, is that there seem to exist data structures that lend themselves to coordination of state evolution in a network of peers without central concurrency control, but that not all data structures have that capability. For example, a simple sequence, with normal {\texttt ins, del} operations, is not capable of that, but a sequence without {\texttt del} and holding more complex nodes that represent a renderable string has that capability. Other examples include grow-only sets, monotonic counters, and many others, including combinations of simpler CRDTs.

CRDTs come in two flavors: state-based and operation-based. They have been proven to be equivalent~\cite{shapiro11}, so we will focus on state-based CRDT. It has been proven~\cite{shapiro11} that a sufficient condition for state-based CRDT to achieve eventual consistency in a network of peers is for the merge method to be commutative, associative, and idempotent, and for state to be monotonically non-decreasing across updates. This is quite a high bar, but it can be met by many data types. Essentially, CRDTs implement the illusion (via rendering) of mutable state with data types that never really forget anything, therefore requiring, in principle, an ever-increasing amount of memory. 

CRDTs have been implemented with great success in industry-grade middleware -- e.g. Redis~\cite{redis}. However, as the scale of the system grows, it takes longer for synchronization to be achieved, because CRDTs require transmission of the full state of the object. Recently, ~\cite{ALMEIDA2018} proposed delta-state CRDT, which are capable of transmitting the state in an incremental manner. This relates to what all version control systems, including GoT, do.

CRDTs are a mathematically sound mechanism to distribute data in cases where eventual consistency of data is needed and where the application programmers can have some control over a few data types. But they are still far from being a general-purpose programming model for replicated objects, which is the goal of our work.

\noindent
{\bf Application-Level Consistency}

\noindent
Recently, there has been some work in exposing data consistency issues to application programmers in a structured manner, in the same line as GoT. Concurrent Revisions~\cite{semantics-of-concurrent-revisions-2} (CR), and subsequent work~\cite{eventually-consistent-transactions, cloud-types-for-eventual-consistency}, introduced the concept of expressing computation as a fork-join automata through revisions. This approach uses versioned variables that are forked between threads or applications, modified, and finally, joined back. Versioning the objects allows for powerful conflict resolution mechanisms that can be specified by the data model, and is not an inherent and inflexible rule of the system. Our conflict resolution functions are similar to CR's, with the difference that our functions are not type-bound, but dataframe-bound. This was one of the most obvious change we had to make, as it became clear that different nodes need to resolve conflicts differently for the same types.

More recently, TARDiS~\cite{tardis} is an asynchronously replicated, multi-master key value store that tries to address some of the problems of ensuring scalable causal consistency. Many of the design choices in GoT are also found in TARDiS and for similar reasons. Neither TARDiS nor GoT abstract the interaction with the data as sequential operations. Both have explicit primitives to observe and resolve concurrency, and distribute changes. Conflict resolutions in TARDiS also include powerful three-way functions that include the original version and provide for meaningful semantic merges. There are also several differences. In GoT, the dataframe is not a database. A database is first and foremost a single service that is shared between multiple applications. In GoT, one dataframe is tied to only one node. Nodes communicate information with each other using a transfer protocol. In a database, data is placed in it for both the application's reference, and for other applications to see. In GoT, data is placed in an external dataframe solely for other nodes to use. Its own dataframe serves as the local context of the node, and serves as the data store for only that node. In this respect it is closer to an object heap than a database. A database like TARDiS expects multiple applications to connect to them. Each application interacting with a TARDiS store is given its own branch of a shared DAG to execute in and merging conflicts becomes a shared task. In GoT, each node is solely responsible for resolving the conflicts in its dataframe.

Another very important difference between TARDiS and GoT is that in GoT, the version graph is stored, edge first, as a sequence of delta changes instead of the node first approach of storing multiple versions of the state. In GoT, complete versions are calculated by merging deltas and are cached for quick lookup in the snapshot. This facilitates very quick patching between dataframes. With full revisions instead of deltas, calculating a deltas for patching is costly. Since the number of entries in a delta is less than or equal to the number of entries in the whole version, it is always faster to apply a delta on a version than it is to create a delta by diff-ing two versions.

%% file: 8_conclusions.tex
\section{Conclusions and Future Work}
\label{sec:conclusions}

In this paper, we present Global Object Tracker (GoT) as a formal object-oriented programming model for object state synchronization among components in a distributed context. A GoT application consists of nodes that are synchronizing over shared data. Each node maintains the revision history of the objects as a graph of delta changes. A local snapshot is provided at each node to provide stable reads, and effective versioning. Updates to the objects can be shared between nodes and their version graphs via delta changes. The model mirrors the processes and interfaces found in decentralized version control systems like Git and brings along three important advantages. First, states can be shared using deltas. This significantly reduces network traffic and improves performance. Second, version graphs allow for powerful conflict detection mechanisms. In GoT, the version graph preserves the causal relations between state changes, and provide an parent version representing the state before the conflicting changes occurred. We complement this powerful conflict detection with the ability to define custom merge strategies to resolve the conflicts. Finally, having a familiar model like Git be the basis of data synchronization is an advantage as developers of the system can have a stronger mental model for reasoning over the operations in GoT.

The pure form of GoT, however, is impractical for real systems. Unchecked version growth can quickly put a strain on resources. Sharing every intermediate version of the version graph on synchronization is wasteful and inefficient. We present solutions to these problems that make GoT feasible in an implementation named \name. Unchecked version growth is aggressively tackled using a reference counting garbage collector that merges multiple deltas that have already been seen into a single large delta. The side effect of having this garbage collector is that nodes have to keep a small amount of the state of the other nodes that interact with it. Intermediate versions of the graph are eliminated before synchronization to make efficient use of network resources. This comes with the trade off that nodes cannot synchronize deltas with nodes that they have never contacted before. This imposes architectural constraints to the network of nodes in \name.

We aim to support the rapid development of multi-agent simulations and distributed competitions for AI bots. We evaluate the feasibility of using \name\ by running experiments using realistic workloads that we expect to see. The experiments show that \name\ is extremely efficient in managing network resources and the aggressive version garbage collector keeps the unbounded memory growth in check. This provides us a good foundation for running distributed AI competitions with a large number of participants.

In the future, we want to explore the use of branches the way Git supports them. We will also explore the possibility of breaking the state into chunks that vary over time. One of the easiest ways to reduce network traffic is to synchronize on only what you want. While GoT allows for versioning over collections of objects, these collections are static. Responding to dynamic queries requires the version graph to be divided which defeats all the performance that they can provide. Further research is needed to find ways around this tradeoff.

%% file: app_A.tex
\section{Complete Listings of Space Race Example}
\label{sec:app_a}

\begin{lstlisting}[language=Python,basicstyle=\small,
label=lst:sr_datamodel_extended, captionpos=b, caption=The data model for the multiplayer Space Race game.]
@pcc_set
class Ship(object):
    oid = primarykey(int); player_id = dimension(str)
    x = dimension(float); y = dimension(float); trips = dimension(int)
    velocity = dimension(float); state = dimension(int)
    def __init__(self, pid, x):
        self.oid = random.randint(0, sys.maxsize)
        ...other initializations, including non-shared fields...
    def go(self):
        self.velocity = -100.0; self.state = ShipState.RUNNING
    def move(self, delta_t):
        if self.state == ShipState.RUNNING:
            self.y += (self.velocity * delta_t)
        if self.y <= 0: # We got to the finish line!
            self.reset(); return True
        return False
    def collision(self):
        self.state = ShipState.DESTROYED; self.velocity = 0.0

@pcc_set
class Asteroid(object):
    oid = primarykey(int); x = dimension(float)
    y = dimension(float); velocity = dimension(float)
    def __init__(self):
        self.oid = random.randint(0, sys.maxsize)
        ...other initializations, including non-shared fields...
    def move(self, delta_t):
        self.x += (self.velocity * delta_t)
        # Did it reach the end?
        if self.x >= World.WORLD_WIDTH or self.x <= 0:
            self.reset()
\end{lstlisting}

\begin{lstlisting}[language=Python,basicstyle=\small, numbers=left,
label=lst:sr_physics_extended, captionpos=b, caption=The Physics simulator node.]
class Game(object):
    PHYSICS_FPS = 20 # per second; for collisions
    DELTA_TIME = float(1)/PHYSICS_FPS

    def __init__(self, df):
        self.dataframe = df
        self.world = World(); self.current_players = {}
        self.init_asteroids()

    def init_asteroids(self):
        for n in range(World.ASTEROID_COUNT):
            self.dataframe.add_one(Asteroid, Asteroid())
        self.dataframe.commit()

    def play(self):
        game_over = False; x = x_gen()
        while not game_over:
            start_t = time.perf_counter()
            
            self.dataframe.checkout()
            players = self.dataframe.read_all(Player)
            for p in players: # Are there new players?
                if p.oid not in self.current_players:
                    self.current_players[p.oid] = p
                    p.ready(next(x), self.dataframe)
            
            self.move_asteroids() # Move all asteroids
            self.move_ships() # Move all ships
            self.detect_collisions() # Deal with collisions

            self.dataframe.commit()

            elapsed_t = time.perf_counter() - start_t
            time.sleep(Game.DELTA_TIME - elapsed_t)

def sr_physics(dataframe):
    game = Game(dataframe); my_print ("READY FOR NEW GAME")
    while True: 
        game.play(); my_print ("GAME OVER"); time.sleep(WAIT_FOR_START)

def main(port):
    node = GotNode(sr_physics, server_port=port, 
                   dataframe="spacerace.got",
                   Types=[Player, Ship, Asteroid])
    node.start()
\end{lstlisting}

\begin{lstlisting}[language=Python,basicstyle=\small, numbers=left,
label=lst:sr_bot_extended, captionpos=b, caption=The Player nodes.]
SYNC_TIME = 0.3 # secs
def bot_driver(dataframe, player_class):
    dataframe.pull()
    my_player = player_class(dataframe)
    dataframe.add_one(Player, my_player)
    dataframe.commit(); dataframe.push()

    my_player.init_world()
    done = False; trips = 0
    while not done:
        start_t = time.perf_counter()

        dataframe.pull()
        survived = my_player.act()
        dataframe.commit(); dataframe.push()

        if not survived:
            # Timeout
            time.sleep(5)
            my_player.reset()
            dataframe.commit(); dataframe.push()
            continue

        if my_player.ship.trips > trips:
            trips = my_player.ship.trips
            my_print("Number of trips: {0}".format(trips))

        elapsed_t = time.perf_counter() - start_t
        sleep_t = SYNC_TIME - elapsed_t
        if sleep_t > 0:
            time.sleep(sleep_t)

def main():
    args = ... # parse command line args
    player = GotNode(bot_driver, dataframe=args.got, 
                     Types=[Player, Asteroid, Ship])
    player.start(get_class(args.player))
\end{lstlisting}

\begin{lstlisting}[language=Python,basicstyle=\small, numbers=left,
label=lst:sr_vis_extended, captionpos=b, caption=The Pygame visualizer class in the viewer nodes.]
class SpaceRaceSprite(pygame.sprite.Sprite):
    def __init__(self, go):
        pygame.sprite.Sprite.__init__(self)  #call Sprite initializer

        self.game_object = go
        self.rect = self.image.get_rect()
        self.rect.left, self.rect.top = [self.game_object.global_x, 
                                         self.game_object.global_y]

class AsteroidSprite(SpaceRaceSprite):
    def __init__(self, go):
        self.image = pygame.image.load("art/asteroid.png")
        SpaceRaceSprite.__init__(self, go)  #call Sprite initializer

    def move(self, delta):
        if self.game_object.global_x <= World.WORLD_WIDTH and \
           self.game_object.global_x >= 0:
            self.game_object.global_x += (self.game_object.velocity * delta)

        self.rect.left, self.rect.top = [self.game_object.global_x, 
                                         self.game_object.global_y]

class ShipSprite(SpaceRaceSprite):
    def __init__(self, go):
        self.images = [pygame.image.load("art/ship.png")]
        self.image_idx = 0
        self.image = self.images[self.image_idx]
        self.anim_counter = 0
        for i in range(1, 16):
            self.images.append(pygame.image.load("art/s-{0}.png".format(i)))
        SpaceRaceSprite.__init__(self, go)  #call Sprite initializer

    def move(self, delta, asteroids):
        if self.game_object.global_y <= World.WORLD_HEIGHT and \ 
           self.game_object.global_y >= 0:
            self.move_delta(delta)
            # Did we collide?
            if self.game_object.state != ShipState.DESTROYED:
                for a in asteroids:
                    if check_collision(a.game_object, 
                                       self.game_object, 
                                       Visualizer.DELTA_TIME+0.02):
                        break

        self.rect.left, self.rect.top = [self.game_object.global_x, 
                                         self.game_object.global_y]

    def move_delta(self, delta):
        if self.game_object.state != ShipState.DESTROYED:
            self.reset()
            self.game_object.global_y += (self.game_object.velocity * delta)
        else:
            self.next_image()

    def next_image(self):
        if self.anim_counter % 4 == 0:
            self.image_idx = min(self.image_idx + 1, len(self.images) - 1)
            self.image = self.images[self.image_idx]
        self.anim_counter += 1

    def reset(self):
        self.image_idx = 0
        self.image = self.images[0]

class TextBar(object):
    def __init__(self, screen, pos, world):
        self.font=pygame.freetype.SysFont(pygame.freetype.get_default_font(),
                                          10)
        self.screen = screen
        self.pos = pos
        self.world = world
        self.update_ticks = 0
        self.message = ""

    def display(self):
        if self.update_ticks % Visualizer.FPS == 0:
            self.message = " " * 20
            trios = [(s.player_id, s.global_x, s.trips) \
                     for s in self.world.ships.values()]
            trios.sort(key=operator.itemgetter(1))
            previous_len = 0
            for name, position, trips in trios:
                nm = clip(name)
                self.message += " " * (int(position/10) - previous_len)
                self.message += nm
                previous_len += len(nm)
        self.update_ticks += 1
        self.font.render_to(self.screen,self.pos,self.message,(0, 255, 0))

class Visualizer(object):
    FPS = 50
    DELTA_TIME = float(1)/FPS
    SYNC_TICKS = 50
    WORLD_Y_OFFSET = 100
    def __init__(self, world):

        pygame.init()
        self.world = world
        self.width = World.WORLD_WIDTH
        self.height = World.WORLD_HEIGHT + Visualizer.WORLD_Y_OFFSET
        self.screen = pygame.display.set_mode((self.width, self.height))
        self.info = TextBar(self.screen, 
                    (0, self.height - Visualizer.WORLD_Y_OFFSET + 10), 
                    world)

        self.listeners = []

        self.asteroids = []
        # Create the asteroid sprites
        for a in world.asteroids.values():
            self.asteroids.append(AsteroidSprite(a))

        self.ships = {}

    def run(self):
        done = False
        while not done:
            start_t = time.perf_counter()

            self.screen.fill((0, 0, 0))
            for event in pygame.event.get():
                if event.type == KEYDOWN and event.key == K_ESCAPE:
                    pygame.quit()
                    done = True
                if event.type == MOUSEBUTTONDOWN or event.type == 5:
                    for l in self.listeners:
                        l.mouse_down(event)

            # Are there new ships?
            if len(self.world.ships) > len(self.ships):
                for id, go in self.world.ships.items():
                    if id not in self.ships:
                        self.ships[id] = ShipSprite(go)

            # Move asteroids
            for obj in self.asteroids:
                obj.move(Visualizer.DELTA_TIME)
                self.screen.blit(obj.image, obj.rect) 
            # Move ships
            for obj in list(self.ships.values()):
                obj.move(Visualizer.DELTA_TIME, self.asteroids)
                self.screen.blit(obj.image, obj.rect) 

            # Text bar
            self.info.display()

            # Let others inject messages
            for l in self.listeners:
                l.update_screen()

            pygame.display.update()

            elapsed_t = time.perf_counter() - start_t
            sleep_t = Visualizer.DELTA_TIME - elapsed_t
            if sleep_t > 0:
                time.sleep(sleep_t)

    def register(self, listener):
        self.listeners.append(listener)

\end{lstlisting}

\newpage
\section{Latency of operations on localhost}
\label{sec:app_b}
\begin{figure}[h]
\centering
\includegraphics[width=\textwidth]{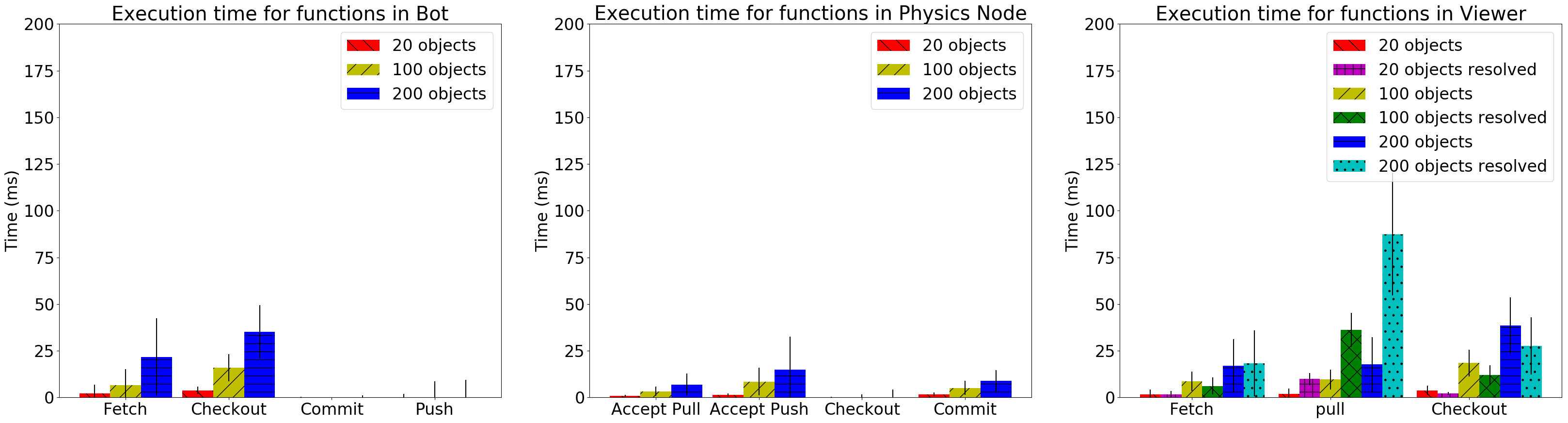}
\caption{Median execution times of multiple functions in each node when in the same machine.}
\label{fig:localhost_bar}
\end{figure}

\newpage